\documentclass{emulateapj}

\usepackage{natbib}
\usepackage{graphicx}

\newcommand{\PH}[1]{\ensuremath{#1_\textrm{\tiny ph}}}
\newcommand{\GJ}[1]{\ensuremath{#1_\textrm{\tiny GJ}}}
\newcommand{\QPO}[1]{\ensuremath{#1_\textrm{\tiny QPO}}}
\newcommand{\TWIST}[1]{\ensuremath{#1_\textrm{\tiny twist}}}
\newcommand{\Pol}[1]{\ensuremath{#1_\mathrm{pol}}}

\newcommand{\dX}{\ensuremath{\Delta{}x}}
\newcommand{\rG}{\ensuremath{r_\mathrm{res}}}
\newcommand{\RNS}{\ensuremath{R_\mathrm{NS}}}

\graphicspath{{./figures/}}

\begin{document}

\title{On the nature of QPO in the tail of SGR giant flares}

\author{A.N.~Timokhin}
\affil{Physics Department, Ben-Gurion
  University of the Negev, Beer-Sheva, Israel\\
Sternberg Astronomical Institute, Moscow, Russia}

\author{D.~Eichler and Yu.~Lyubarsky}
\affil{Physics Department, Ben-Gurion University of the Negev, Beer-Sheva, Israel}

\begin{abstract}
  A model is presented for the quasiperiodic component of magnetar
  emission during the tail phase of giant flares. The model invokes
  modulation of the particle number density in the magnetosphere.  The
  magnetospheric currents are modulated by torsional motion of the
  surface and we calculate that the amplitude of neutron star surface
  oscillation should be $\sim{}1\%$ of the NS radius in order to
  produce the observed features in the power spectrum.  Using an
  axisymmetric analytical model for structure of the magnetosphere of
  an oscillating NS , we calculate the angular distribution of the
  optical depth to the resonant Compton scattering.  The anisotropy of
  the optical depth may be why QPO are observed only at particular
  rotational phases.
\end{abstract}

\keywords{stars:magnetic field -- stars:neutron -- stars:oscillations}

\section{Introduction}
\label{sec:introduction}

\citet{Duncan1998} proposed that a magnetar undergoing a giant flare
would be set into oscillations.  \citet{Timokhin2000,Timokhin2006}
proposed that torsional oscillations could modulate the local
Goldreich-Julian charge density and noted that changes in the
Goldreich-Julian charge density are proportional to the harmonic
numbers $l,m$.  \citet{Israel2005} and \citet{Strohmayer2005} have
reported detections of quasi-periodical features in the power spectra
of SGR 1806-20 and 1900+40, which they interpret to be n=0 high-$l$
torsional oscillations of the crust. These features in the power
spectra were detected mostly in hard (probably non-thermal) X-rays in
the energy range up to 200 keV; the frequencies of this QPO range from
18 Hz up to 600 Hz.  The QPO emission is observed only during certain
rotational phases.  One of several remarkable features of this
discovery is the rather large amplitude of the luminosity modulation,
of the order of 10\% for the high $l$ modes, given that a crust would
break or melt if the crustal oscillations were too violent. The shear
strain of crustal oscillations is probably limited to about
$\sim10\%$, which, for $l\sim10$, would result in a limit for the
oscillation amplitude about one or two percent of the NS radius
$\RNS$, unless vortex pinning and/or strong magnetic field make the
crust more resilient.

Even more remarkable is that such large modulation appears most
prominently even at certain high $l$ harmonics. Even if temperature
fluctuations of order several percent could be somehow achieved at the
photosphere, it would not lead to such large luminosity modulation if
many maxima and minima were visible to the observer at any given time.
Moreover, the high value of $l$ suggests that energy in crustal
oscillations is triggered from a relatively small part of the crust's
surface and is distributed into a large number of modes.  It also
raises the possibility that the deep crustal heating that can account
for the medium and long term X-ray afterglow following giant flares
and other periods of magnetic activity
\citep{Lyubarsky2002,Kouveliotou2003} may be powered by crustal
oscillations.

In this paper we propose a model for the amplitude variations that
appears to support the idea that the oscillations are indeed torsional
oscillations of the crust, and may also support the suggestions
\citet{Thompson2002} that the magnetospheric currents play a
significant role in the magnetar emission at high photon energy.

\section{The model}
\label{sec:model}

We suggest that the modulation of the hard X-ray radiation visible in
the tail of the giant flares is caused by modulation of the particle
number density in the magnetosphere due to shaking of the magnetic
field lines by oscillations of the NS crust.  The specific candidate
emission mechanisms can be cyclotron scattering of thermal photons
\citep{Lyutikov2006}, non-resonant scattering of thermal photons, or
bremsstrahlung off a target corona, though a detailed investigation of
the latter two is not attempted here, as they seem to us less
promising.  The electrical resistance created by photon drag during
the tail phase of a giant flare is several orders of magnitude greater
- as is the non-QPO photon background against which the QPO component
must stand out - than during the quiet time persistent emission.  This
consideration leads us to favor resonant scattering, for which the
optical depth offered by the current carriers to the thermal photons
is the greatest%
\footnote{The coronal density may itself be enhanced during tail-phase
  emission by radiative trapping of plasma on closed field lines, and
  this might also be a source of enhanced electrical resistivity that
  directly taps the energy in magnetic field twist. However, here we
  focus only on resonant scattering.}.

In order to support the twist of the magnetic field, a current density
$j$ must flow along magnetic field lines. The minimum particle density
in the magnetosphere of the magnetar is therefore $j/(ec)$. The
strength of the magnetic field where resonant inverse Compton
scattering on electrons occurs is, in the absence of Doppler shift,
\begin{equation}
  \label{eq:Bres}
  B_\mathrm{res} = 0.086 \PH{E}^\mathrm{[keV]} 10^{12} \mathrm{G}\,,
\end{equation}
where $\PH{E}^\mathrm{[keV]}$ is the initial energy of the scattered
photons in units of keV.  For magnetars with the surface magnetic
field $10^{15}-10^{14}$~G the resonant surface for hard X-rays lies at
the distance of the order of $\sim{}10$ stellar radii ($\RNS$) or
less.  For the observed frequencies of the QPO the electromagnetic
wavelength $\QPO{\lambda}=c/\QPO{\nu}$ corresponding to these
oscillations are of the order 50 to $1700~\RNS$.  Hence the resonant
surface is well within the wave zone and the magnetosphere can be
considered as quasistatic.  All physical quantities changes with time
as $e^{-i\omega{}t}$.

If the reconfiguration of the magnetic field induces oscillations of
the NS crust \citep{Duncan1998}, the motion of the star surface moves
the magnetic field lines frozen into the surface.  The motion of the
field lines induces the electric field and, for modes which move the
footpoints of a given magnetic field line in opposite directions,
induces a twist in the field line.  The twist of the field lines
induces electric current along the lines and hence changes the
particle number density in the magnetosphere. As we show in
appendix~\ref{sec:appendix_A} only modes which move the footpoints of
a given magnetic field line in opposite directions and thus induce the
electric current along magnetic field lines could modulate hard
X-radiation of magnetar.

The optical depth to the resonant Compton scattering on particle with
the charge $Ze$ is (cf. eq.~(23) in \citet{Thompson2002})
\begin{equation}
  \label{eq:tau_exact}
  \tau(\theta)=\pi^2 Ze n\;
  (1+\cos^2\theta_{kB})\left|\frac{dB}{dr}\right|^{-1}
  \,,
\end{equation}
where all quantities are taken at the corresponding point of the
resonant surface $(\rG,\theta)$; $n$ is the resonant particle number
density, $\theta_{kB}$ is the angle between the photon wave vector and
the magnetic field.  For dipolar magnetic field
$|dB/dr|\simeq(3/2)B/r$.  The current density induced by the twist of
the magnetic field line is
\begin{equation}
  \label{eq:jTwist}
  \TWIST{j}=\frac{c}{4\pi}\nabla\times\vec{B} \sim
  \frac{c}{4\pi} \frac{\delta{}B}{\dX} \sim
  c\frac1{2} \frac{\xi}{\dX} \frac{B}{\pi{}s},
\end{equation}
where $\delta{}B\sim{}2(\xi/s){}B$ is the variation of the magnetic
field due to twist of the magnetic field line; $\dX$ is the
characteristic scale of the electromagnetic field variation; $s$ is
the length of the corresponding magnetic field line; $\xi$ is the
oscillation amplitude.  Assuming that the plasma density is
proportional to the current density, see eq.~(\ref{n_trough_kappa}) in
appendix~\ref{sec:appendix_B}, we get for the optical depth
\begin{equation}
  \label{eq:tau_estimation}
  \tau \sim
  \pi^2 \kappa \frac{\TWIST{j}}{c} \frac{2}{3} \frac{\rG}{B} \sim
  \frac{\pi}{3} \kappa \frac{\xi}{\dX} \frac{\rG}{s}
\end{equation}
where $\kappa$ is multiplicity of pair plasma.  The maximum optical
depth is achieved near the equator, where $s/\rG\simeq2.6$; for
oscillation mode with the harmonic number $l$ $\dX\simeq\pi\RNS/l$.
Using these $\tau$ can be estimated as
\begin{equation}
  \label{eq:tau_Approx}
  \tau
  \sim \kappa l\frac{\xi}{3\RNS} \frac{\rG}{s}
  \sim \kappa \frac{l}{8} \frac{\xi}{\RNS}
  \,.
\end{equation}
Hence the modulation of the optical depth of the order of $0.01\kappa$
would require oscillation amplitude for the harmonic number
$l\simeq10$ of the order of 1\%.  Such oscillation would produce shear
strain of the NS crust of the order of 10\%, implying that the nearest
neighbor distance oscillates by about 1\% (In
appendix~\ref{sec:appendix_B} we argue that $\kappa$ may be of the
order of a few or larger, which would provide optical depth modulation
of the order of $\sim10\%$).  The question is whether such a crustal
strain could accommodate the observed QPO in the flux.  We now do a
more quantitative calculation relating optical depth to crustal strain
and flux modulation.

\section{Axisymmetric force-free magnetosphere of an oscillating
  neutron star with dipole magnetic field}
\label{sec:axisymm-force-free}

Consider an idealized case of a non-rotating star with dipole magnetic
field oscillating in toroidal oscillation mode with small amplitude.
We consider disturbance of the magnetosphere by toroidal modes with
even value of $l$ and $m=0$.  We use a spherical coordinate system
centered at the NS center, with $z$-axis parallel to the dipole
momentum and the same representation for magnetic field and currents
in the force-free magnetosphere as in \citet{Timokhin2006a}.  For an
axisymmetric force-free field, the toroidal component induced by the
oscillations is
\begin{equation}
  \label{eq:Bphi_I}
  B_\phi =\frac{4\pi}{c}\frac{I}{r\sin\theta}
  \,,
\end{equation}
where $2\pi{}I$ is the total induced poloidal current flowing between
the pole and the colatitude $\theta$.  For small perturbation of the
magnetic field we have
\begin{equation}
  \label{eq:df_ds}
  \frac{r\sin\theta\:d\phi}{ds} = \frac{B_\phi}{\Pol{B}}
  \,,
\end{equation}
where $\Pol{B}$ is the poloidal component of the magnetic field and
$ds$ is the distance along the field line.  The total twist of the
magnetic field line (the difference $\phi_1-\phi_2$ of the field line
footpoint azimuthal angles) $\delta\phi_*$ is
\begin{equation}
  \label{eq:delta_phi_tot}
  \delta\phi_*=\int^{\pi-\theta_*}_{\theta_*}\frac{d\phi}{d\theta}\,d\theta
  \,,
\end{equation}
where $\theta_*$ is the colatitude of the magnetic field line
footpoint in the upper hemisphere.  The derivative $d\phi/d\theta$ can
be expressed through the derivative $d\phi/ds$ as
\begin{equation}
  \label{eq:df_dq_through_df_ds}
  \frac{d\phi}{d\theta} = \frac{d\phi}{ds}\frac{ds}{d\theta} =
  \frac{d\phi}{ds} \RNS \frac{\sin\theta}{\sin^2\theta_*}
  \sqrt{1+3\cos^2\theta}
  \,,
\end{equation}
and $d\phi/ds$ can be obtained from eq.(\ref{eq:df_ds}).  Expressing
$B_\phi$ through the poloidal current $I$, eq.~(\ref{eq:Bphi_I}), and
performing the integration, we get for the total induced poloidal
current
\begin{equation}
  \label{eq:I_tot}
  I=\frac1{2}\delta\phi_*\frac{c}{8\pi}B_0\RNS
  \frac{\sin^4\theta_*}{\cos\theta_*}
  \,.
\end{equation}
The total twist of the magnetic field line $\delta\phi_*$ is
\begin{equation}
  \label{eq:df_total}
  \delta\phi_*(\theta_*) = 2\frac{\xi(\theta_*)}{\RNS\,\sin\theta_*}
  \,,
\end{equation}
where $\xi(\theta_*)$ is the displacement of the magnetic field line
footpoint.  For the toroidal oscillation modes under consideration,
the displacement of the NS surface $\xi(\theta_*)$ is
\begin{equation}
  \label{eq:xi_toroidal}
  \xi(\theta_*) = -\xi_0 \frac{dY_{l0}(\theta_*)}{d\theta_*}
  \,,
\end{equation}
where $\xi_0$ is the mode amplitude and $Y_{l0}$ is the spherical
harmonic of order $l,0$. The current density can be expressed through
the magnetic flux function $\psi$ as
\begin{equation}
  \label{eq:j_psi}
  j=\frac{\nabla{}I\times\vec{e}_\phi}{r\sin\theta}=\frac{dI}{d\psi}\Pol{\vec{B}}
\end{equation}
The magnetic flux function $\psi$ for the dipole field is
\begin{equation}
  \label{eq:psi_dipole}
  \psi=\frac{B_0\RNS^3}{2}\frac{\sin^2\theta}{r}
  \,,
\end{equation}
where $B_0$ is magnetic field strength at the pole.  The derivative
$dI/d\psi$ is
\begin{equation}
  \label{eq:dI_dpsi}
  \frac{dI}{d\psi}=\frac{dI}{d\theta_*}\frac{d\theta_*}{d\psi}=
  \frac{dI}{d\theta_*}\frac{1}{B_0\RNS^2\sin\theta_*\cos\theta_*}
  \,.
\end{equation}
Combining equations~(\ref{eq:I_tot})-(\ref{eq:dI_dpsi}) we obtain
\begin{equation}
  \label{eq:j_final}
  j=j_0 \left(\frac{r}{\RNS}\right)^{-3}
  \sqrt{1+3\cos^2\theta}\: A_l[\theta_*(\theta,r)]
  \,,
\end{equation}
where
\begin{eqnarray}
  \label{eq:j_final__j_0}
  j_0 & = & \frac{c}{16\pi} \frac{B_0}{\RNS} \frac{\xi_0}{\RNS}\\
  \label{eq:j_final__A_l}
  A_l & = &
  \frac{\sin\theta_*}{\cos^2\theta_*}
  \left(
    \frac{1+2\cos^2\theta_*}{\cos\theta_*}\frac{dY_{l0}}{d\theta_*} +
    \sin\theta_* \frac{d^2Y_{l0}}{d\theta_*^2}
  \right)
\end{eqnarray}

\begin{figure}
  \includegraphics[width=\columnwidth]{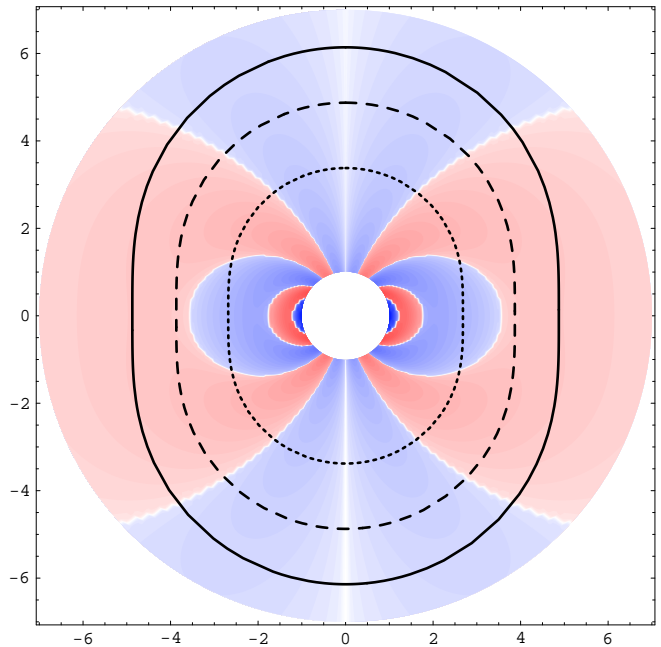}
  \caption{Current density distribution in the magnetosphere of a NS
    with dipole magnetic field oscillating in toroidal mode
    $(l=10,m=0)$.  The intensity of color in each point is
    proportional to $|j(r,\theta)|^{1/6}$ (this moderates contrast of
    the plot).  Positive values of the current density are shown by
    red colors, negative ones by the blue color.  By the black lines
    the resonant surfaces are shown for three photon energies
    $\PH{E}=5;10;30$~keV -- solid, dashed and dotted lines
    correspondingly.  The magnetic field at the pole is assumed to be
    $B_0=10^{14}$~G.
    \label{fig:j10}}
\end{figure}

As an example, we plot in Fig.~\ref{fig:j10} the current density
distribution for toroidal mode $l=10$.  The current number density is
strongly modulated in latitudinal direction and therefore the resonant
Compton optical depth is highly anisotropic.

\begin{figure*}
  \begin{center}
    \includegraphics[width=\textwidth,clip]{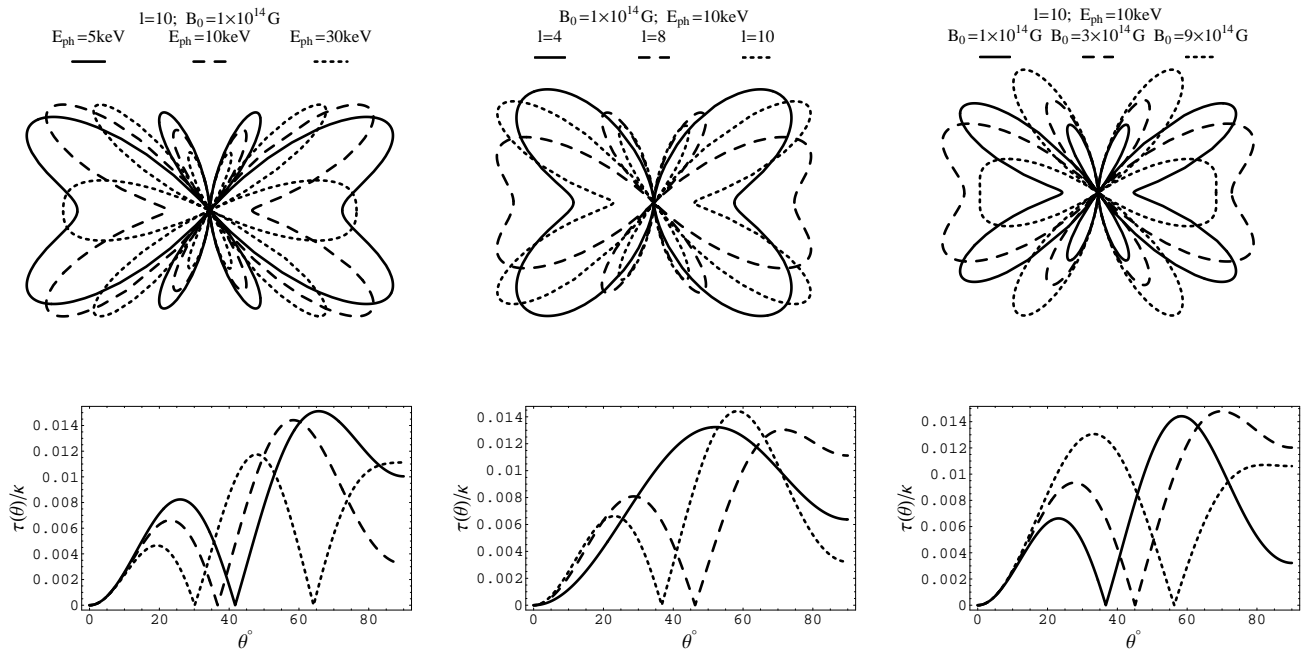}
  \end{center}
  \caption{Angular distribution of the optical depths to resonant
    Compton scattering for radially moving photons in the
    magnetosphere of an oscillating NS for oscillations with the
    amplitude $\xi_0=0.01\RNS$.  In the upper figures the angular
    distribution $\tau(\theta)/\kappa$ is shown in polar coordinates
    $(\tau(\theta)/\kappa,\theta)$, in lower figures the same
    distribution is shown as a plot $\tau(\theta)$ for
    $\theta\in[0,90^\circ]$.  In the left figures the optical depth is
    shown for 3 different initial energies of the scattered photons
    ($\PH{E}=5;10;30$~keV), the magnetic field $B_0=10^{14}$~G and the
    harmonic number of the toroidal mode $l=10$ are fixed.  In the
    middle figures the optical depth is shown for 3 different modes
    ($l=4;6;10$), the magnetic field $B_0=10^{14}$~G and the initial
    energy of the scattered photons $\PH{E}=10$~keV are fixed.  In the
    right figures the optical depth is shown for 3 different values of
    the NS magnetic field ($B_0=1;3;9\times10^{14}$~G), the harmonic
    number of the toroidal mode $l=10$ and the initial energy of the
    scattered photons $\PH{E}=10$~keV are fixed.
    \label{fig:tau_theta}}
\end{figure*}

Now we can obtain an expression for the optical depth.  Substituting
the expression for the current, eq.(\ref{eq:j_final}), into the
formula for the optical depth~(\ref{eq:tau_exact}), we get
\begin{equation}
  \label{eq:tau_twist_exact}
  \tau =
  \kappa
  \frac{\pi}{24} \frac{\xi_0}{\RNS} \rG
  \frac{1+7\cos^2\theta}{1+3\cos^2\theta} A_l[\theta_*(\theta,r)]
  \,,
\end{equation}
Here for simplicity, as in \citet{Thompson2002}, we assume radially
streaming photons.

\begin{figure}
  \includegraphics[width=\columnwidth]{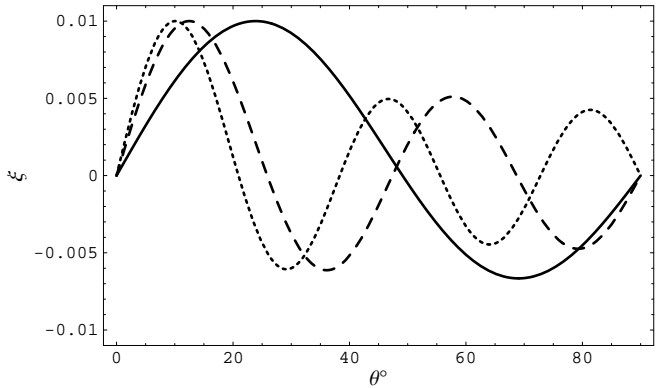}
  \caption{Displacement amplitude of the NS surface normalized to the
    NS radius as a function of the colatitude for toroidal oscillation
    modes $l=4$ -- solid line; $l=8$ -- dashed line; $l=10$ -- dotted
    line.  The mode amplitude $\xi_0=0.01\RNS$.}
   \label{fig:xi}
\end{figure}

\begin{figure}
  \includegraphics[width=\columnwidth]{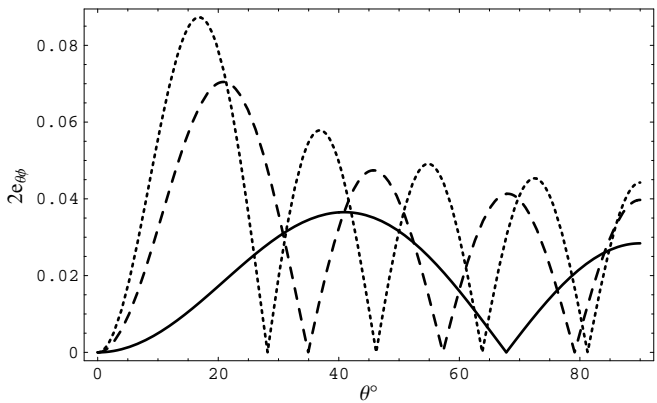}
  \caption{The largest component of the strain tensor,
    $2e_{\theta\phi}$, as a function of the colatitude for toroidal
    oscillation modes $l=4$ -- solid line; $l=8$ -- dashed line;
    $l=10$ -- dotted line.  The mode amplitude $\xi_0=0.01\RNS$.}
   \label{fig:strain}
\end{figure}

If the resonant surface is located not very far from the NS, the
optical depth has several maxima at different colatitudes.  The
position of these maxima as well as their amplitude will depend on the
shape and position of the resonant surface.  The latter depends on the
magnetic field strength as well as on the energy of the upscattered
photons.  In Fig.~\ref{fig:j10} we plot three resonant surfaces for
different energies of the upscattered photons for the polar magnetic
field $B_0=10^{14}$~G.  It is evident that for different photon
energies the angular distribution of the optical depth will be
different.  In Fig.~\ref{fig:tau_theta} we plot the angular
distribution of the optical depth for different energies of the
upscattered photons, different oscillation modes and different
magnetic field strength.  In each of these plots we assumed the
amplitude of the oscillation mode $\xi_0=0.01\RNS$. Latitudinal
distribution of the oscillation amplitude $\xi(\theta_*)$ as well as
the largest component of the strain tensor $e_{\theta\phi}$ \citep[see
e.g.][]{LandauLifshits7} for the considered modes $l=4;8;10$ are shown
in Figs.~\ref{fig:xi} and \ref{fig:strain} correspondingly.

The optical depth distribution has the following properties
\begin{itemize}

\item The latitude of peak optical depth is strongly energy dependent
  for the low latitude peaks and less so for the peaks near the polar
  axis (see Fig.~\ref{fig:j10}).

\item The maximum optical depth does not depend significantly on the
  initial photon energy (Fig.~\ref{fig:tau_theta},~left panels).

\item For moderate l ($l\lesssim{}10$) the optical depth produced by
  mode amplitude $\xi_0$ varies little with $l$, although the angular
  dependence is naturally different, see Fig.~\ref{fig:tau_theta}
  (middle panels). For larger $l$ equation~(\ref{eq:tau_Approx}) gives
  a rather good estimate for the maximum value of the optical depth,
  which, for the same mode amplitude at the magnetic field line
  footpoint $\xi(\theta_*)$, grows with increasing $l$.

\item The maximum value of the optical depth does not depend
  significantly on the strength of the magnetar magnetic field for
  commonly accepted parameters of $B_0=10^{14}-10^{15}$~G, see
  Fig.~\ref{fig:tau_theta}, right panels.

\end{itemize}

We consider now three possibilities for modulation of hard
X-radiation: a) modulation is independent on photon energy change and
depends only on anisotropic obscuration of photons generated within
the resonant surface; b) modulation is achieved by upscattering of
photons in energy by either (i) non-relativistic electrons or (ii) at
least moderately relativistic electrons.

Case a) requires an anisotropic optical depth of the order of 10\%.
This is easily achieved with the multiplicity $\kappa$ of the order of
10, such multiplicity is easily achieved because pair-production
requires primary electron energies of the order of $\simeq{}10mc^2$,
where $m$ is the electron mass, (see appendix~\ref{sec:appendix_B}).

For the case b(i) the fluctuation of the photon flux is increased by
the Compton-Getting effect as
\begin{equation}
  \label{eq:deltaF_F_bi}
  \frac{\Delta{}\PH{F}}{\PH{F}} =
  \tau\frac{U}{c}
  \left[-\frac{d\log{}\PH{F}(\PH{E})}{d\log{}\PH{E}} \right] \approx
  4 \tau\frac{U}{c}
  \,,
\end{equation}
where $U$ is the characteristic electron velocity at the scattering
point (the sum of bulk, thermal and drift velocity) and
$F_\mathrm{ph}$ is the number density of photons with the energy
$\PH{E}$.  We have used the time integrated photon spectrum
$F_\mathrm{ph}(\PH{E})$ of the tail phase reported by
\citet{Hurley2005}.  This is more than the contribution of the case a)
if $U/c>1/4$. It gives at least 6\% modulation of the hard X-ray flux
for displacement amplitude of the order of $0.01\RNS$.

Case b(ii) minimizes the optical depth implied by the twisted field,
but, on the other hand, it enables the scattering process to lift
photons directly from the thermal peak at $\PH{E}\simeq10~keV$ to the
high energy tail.  We estimate that the unmodulated tail represents
about 0.1 of the total photon number \citep[see][]{Hurley2005}, so a
QPO modulation amplitude in the high energy tail of 10\% would require
an optical depth to resonant inverse Compton scattering of about 1\%.

For oscillation modes with moderate $l<10$ the maximum value of
$\tau/\kappa$ reaches $\sim1\%~(1.5)\%$ when the displacement of the
NS crust is $0.6\%~(1\%)$, see Fig.\ref{fig:xi}.  The strain of the
crust caused by a particular mode does not exceed 10\%, see
Fig.\ref{fig:strain}.  Not surprisingly, this is close to the maximum
beyond which the crust would either break or plastically deform, but
it seems tolerable, unless there are so many modes excited that the
root mean square strain is much more than per individual mode. The
strong magnetic field may provide a rubbery consistency to the crust
material so that it could withstand larger reversible strain than
brittle crystal, and this would be an implication of the likely
possibility that many other modes were excited in addition to those
that produced observable QPO behavior.

Angular modulation of the optical depth to the resonant Compton
scattering may result in temporal modulation of the hard X-ray
photons.  Indeed the QPO features are observed only at specific phases
of the X-ray light curve \citep{Strohmayer2006,Israel2005}. A
quantitative fit to the light curves is beyond the scope of this
article, but we suspect that angular dependence of the optical depth
is the main reason that the QPO features are rotational phase
dependent.  Altogether, the proposed scenario seems to be in good
agreement with the interpretation of the largest amplitude ($\sim
10\%$) QPOs as somehow resulting from NS oscillations with harmonic
number $l\lesssim{}10$ \citep[e.g.][]{Strohmayer2006,Israel2005}.

\section{Discussion}
\label{sec:discussion}

In all of the above, we have assumed for simplicity that the torsional
oscillations and magnetic field are coaxial and axisymmetric. In
reality, a more complex topology of the magnetic field is likely,
hence a more complicated $l$ dependence.  The general principle
nevertheless remains that some modes twist field line more than others
and that some field lines will be twisted more than others for any
given mode.

Our proposal that the QPO oscillations result from oscillation-induced
enhancements of the magnetospheric currents is based on the premise
that the emission or upscattering increases with density.  This would
not work, for example, for blackbody emission. We therefore considered
it significant that these QPO were observed in the hard X-ray band and
not in the thermal peak.  Our model predicts that QPO behavior should
be significantly less in any black body thermal component.  Similarly
the fact that QPO are seen at high photon energies but not at low
photon energies favors upscattering over anisotropic obscuration.  Our
interpretation of the QPO peaks in the tail phase emission of the
giant flares seems consistent with the overall picture of magnetar
emission that has emerged over the past several years. In particular
it is consistent with the view that the hard X-radiation is due to
non-thermal reprocessing by the magnetospheric current carriers or
direct emission from a hot corona. It is also consistent with the
interpretation that these oscillations are torsional rather than some
other kind.

One unsurprising implication of the remarkably large amplitude of the
QPO modulation is that the crust receives an enormous amount of
mechanical energy from the giant flare. We estimate over $10^{42}$ erg
per mode if the amplitude of oscillation is $10^{-2}$ neutron star
radii and the QPO frequency centered on 150 Hz. Assuming that the the
modes are indeed torsional modes, it follows that their mechanical
energy deposition extends throughout the entire crust. The fact that
the measured Q of these oscillations is always much less than the
number of periods that the QPO behavior lasts is quite curious, and
the most obvious interpretations are either that a) the oscillations
damp on a short time compared to the overall duration, requiring
continuous excitation, b) many modes are excited ($\gg 10$, as at
least 10 separate frequencies are detected in the individual source)
with an unresolved spreads in frequencies clustering around the
observed central ones.  Either interpretation implies that far more
than $10^{43}$ ergs are excited.  we may suppose that over
$10^{45}$~erg is released in the inner crust as mechanical motion.
This is far more than the emission during the tail phase and probably
goes mostly into heating the crust.  This is consistent with the
assumption that over $10^{45}$ erg can be deposited in the deep crust
that was made \citep{Kouveliotou2003,Eichler2006} to account for
observed long term X-ray afterglow from magnetars 1627-41 and 1900+14.

In each case a decent fit to the observed cooling curve requires a
large enough mass of the NS that the direct Urca process cools the
core (at least $1.4 M_\odot$).  Combining this result with analysis
shown in Fig.~5 in \citet{Strohmayer2005} would set strong limits on
the equation of state of the NS.

\acknowledgments

We are grateful to A.~Watts for fruitful discussions.  This work was
partially supported by the United States - Israel Binational Science
Foundation and by an Israel Science Foundation Center of Excellence
Grant.  ANT also acknowledges financial support from Russian grants
N.Sh.-5218.2006.2, RNP.2.1.1.5940, N.Sh.-10181.2006.2. YL also
acknowledges financial support from the German-Israeli Foundation for
Scientific Research and Development.

\bibliography{magnetar}

\begin{thebibliography}{16}
\expandafter\ifx\csname natexlab\endcsname\relax\def\natexlab#1{#1}\fi

\bibitem[{{Beloborodov} \& {Thompson}(2007)}]{Beloborodov2007}
{Beloborodov}, A.~M., \& {Thompson}, C. 2007, \apj, 657, 967

\bibitem[{{Duncan}(1998)}]{Duncan1998}
{Duncan}, R.~C. 1998, \apjl, 498, L45+

\bibitem[{{Eichler} {et~al.}(2006){Eichler}, {Lyubarsky}, {Kouveliotou}, \&
  {Wilson}}]{Eichler2006}
{Eichler}, D., {Lyubarsky}, Y., {Kouveliotou}, C., \& {Wilson}, C.~A. 2006,
  arXiv:astro-ph/0611747v1

\bibitem[{{Hurley} {et~al.}(2005){Hurley}, {Boggs}, {Smith}, {Duncan}, {Lin},
  {Zoglauer}, {Krucker}, {Hurford}, {Hudson}, {Wigger}, {Hajdas}, {Thompson},
  {Mitrofanov}, {Sanin}, {Boynton}, {Fellows}, {von Kienlin}, {Lichti}, {Rau},
  \& {Cline}}]{Hurley2005}
{Hurley}, K., {Boggs}, S.~E., {Smith}, D.~M., {Duncan}, R.~C., {Lin}, R.,
  {Zoglauer}, A., {Krucker}, S., {Hurford}, G., {Hudson}, H., {Wigger}, C.,
  {Hajdas}, W., {Thompson}, C., {Mitrofanov}, I., {Sanin}, A., {Boynton}, W.,
  {Fellows}, C., {von Kienlin}, A., {Lichti}, G., {Rau}, A., \& {Cline}, T.
  2005, \nat, 434, 1098

\bibitem[{{Israel} {et~al.}(2005){Israel}, {Belloni}, {Stella}, {Rephaeli},
  {Gruber}, {Casella}, {Dall'Osso}, {Rea}, {Persic}, \&
  {Rothschild}}]{Israel2005}
{Israel}, G.~L., {Belloni}, T., {Stella}, L., {Rephaeli}, Y., {Gruber}, D.~E.,
  {Casella}, P., {Dall'Osso}, S., {Rea}, N., {Persic}, M., \& {Rothschild},
  R.~E. 2005, \apjl, 628, L53

\bibitem[{Kosevich {et~al.}(1986)Kosevich, Lifshitz, Landau, \&
  Pitaevskii}]{LandauLifshits7}
Kosevich, A.~M., Lifshitz, E.~M., Landau, L.~D., \& Pitaevskii, L.~P. 1986,
  Theoretical Physics, Vol.~7, Theory of Elasticity (Butterworth-Heinemann)

\bibitem[{{Kouveliotou} {et~al.}(2003){Kouveliotou}, {Eichler}, {Woods},
  {Lyubarsky}, {Patel}, {G{\"o}{\u g}{\"u}{\c s}}, {van der Klis}, {Tennant},
  {Wachter}, \& {Hurley}}]{Kouveliotou2003}
{Kouveliotou}, C., {Eichler}, D., {Woods}, P.~M., {Lyubarsky}, Y., {Patel},
  S.~K., {G{\"o}{\u g}{\"u}{\c s}}, E., {van der Klis}, M., {Tennant}, A.,
  {Wachter}, S., \& {Hurley}, K. 2003, \apjl, 596, L79

\bibitem[{{Lyubarsky} {et~al.}(2002){Lyubarsky}, {Eichler}, \&
  {Thompson}}]{Lyubarsky2002}
{Lyubarsky}, Y., {Eichler}, D., \& {Thompson}, C. 2002, \apjl, 580, L69

\bibitem[{{Lyutikov} \& {Gavriil}(2006)}]{Lyutikov2006}
{Lyutikov}, M., \& {Gavriil}, F.~P. 2006, \mnras, 368, 690

\bibitem[{{Strohmayer} \& {Watts}(2005)}]{Strohmayer2005}
{Strohmayer}, T.~E., \& {Watts}, A.~L. 2005, \apjl, 632, L111

\bibitem[{{Strohmayer} \& {Watts}(2006)}]{Strohmayer2006}
---. 2006, \apj, 653, 593

\bibitem[{{Thompson} \& {Duncan}(1995)}]{Thompson1995}
{Thompson}, C., \& {Duncan}, R.~C. 1995, \mnras, 275, 255

\bibitem[{{Thompson} {et~al.}(2002){Thompson}, {Lyutikov}, \&
  {Kulkarni}}]{Thompson2002}
{Thompson}, C., {Lyutikov}, M., \& {Kulkarni}, S.~R. 2002, \apj, 574, 332

\bibitem[{{Timokhin}(2006)}]{Timokhin2006a}
{Timokhin}, A.~N. 2006, \mnras, 368, 1055

\bibitem[{{Timokhin}(2007)}]{Timokhin2006}
---. 2007, \apss, 308, 345

\bibitem[{{Timokhin} {et~al.}(2000){Timokhin}, {Bisnovatyi-Kogan}, \&
  {Spruit}}]{Timokhin2000}
{Timokhin}, A.~N., {Bisnovatyi-Kogan}, G.~S., \& {Spruit}, H.~C. 2000, \mnras,
  316, 734

\end{thebibliography}

\appendix

\section{A. Goldreich-Julian charge density}
\label{sec:appendix_A}

The twist of magnetic field lines induces currents flowing along the
field lines.  Additionally, motion of the field lines induces an
electric field, and, in order to keep the magnetosphere force-free the
charge density in the magnetosphere must contain an oscillating
component - the "oscillational" Goldreich-Julian (GJ) charge density.
Let us compare minimal particle number densities required by changes
in the GJ charge density and by current supporting twist of the
magnetic field lines.

The oscillational GJ charge density could be estimated as
\begin{equation}
  \label{eq:rhoGJ}
  \GJ{\rho} = \frac1{4\pi}\nabla\cdot\vec{E} \sim
  \frac1{4\pi}\frac{v}{c} \frac{B}{\dX} \sim
  \frac1{2}\frac{\xi}{\dX} \frac{B}{\QPO{\lambda}}
\end{equation}
Here $\dX$ is the characteristic scale of the electromagnetic field
variation, $v$ is the amplitude of the oscillational velocity, $\xi$
is the oscillation amplitude.  The current density induced by the
twist of the magnetic field line is given by eq.~(\ref{eq:jTwist}).
The ratio of the corresponding particle number densities is thus
\begin{equation}
  \label{eq:nGJ2nTwist}
  \frac{\TWIST{n}}{\GJ{n}}\equiv
  \frac{\TWIST{j}}{c}\frac1{\GJ{\rho}}\sim
  \frac{\QPO{\lambda}}{\pi{}s} > 1
  \,,
\end{equation}
the last inequality follows from estimates of the sizes of the
resonant surface and the wave zone given in section~\ref{sec:model}.
We note that this estimate is valid for any point in the
magnetosphere, also for a point at the resonant surface.  $s$ is then
the length of the magnetic field line from the considered point to the
corresponding symmetrical point at the resonant surface.  Taking into
account that the resonant surface is at distances of the order of
$\sim4-10\RNS$ we conclude that the particle number density modulation
due to changes in the GJ charge density is always less than the
modulation caused by twist of the magnetic field lines, even for
oscillation with frequencies of the order of 1~kHz, when the wavezone
size is $30\RNS$.

The magnetic field lines get twisted only if their footpoints are
moving in opposite directions.  Hence, in an axisymmetric background
magnetic field only half the oscillation modes twist the field lines.
The current induced by the twist flows along the field line and
changes the particle density in proportion to the local value of the
magnetic field, so the current density with the distance from the NS
would decrease as $\sim{}r^{-3}$. These are the modes, we propose,
which produce observable quasiperiodic modifications in the hard
X-radiation in the tail of the giant flares.

The modes with different symmetry, when the field line footpoints move
in the same directions produces only modulation of the particle number
density due to changes in the GJ charge density.  Such situation was
considered in \citet{Timokhin2000,Timokhin2006}, where analytical 3D
model of the force-free magnetosphere of an oscillating NS was
constructed under assumption that the current along magnetic field
lines induced by the stellar oscillations is negligibly small. In that
case only the \emph{charge} density near the NS is modulated, which as
we shown above introduces much smaller modulation to the particle
number density comparing to the modes twisting magnetic field lines
with the same oscillation amplitude.  Moreover, far from the star, near
the resonant surface, the alteration of the particle number density is
very weak because $\GJ{\rho}$ falls very rapidly with the distance
from the NS, as $r^{-(2l+1)}$ for a mode with harmonic number $l$.  The
changes in the particle number density near the resonant surface, in
that case, is negligibly small and such modes will not produce
detectable modulation of the hard X-ray emission.

\section{B. Plasma production in the magnetosphere}
\label{sec:appendix_B}

At the tail phase of the giant burst, a magnetically trapped fireball
radiates the energy stored during the initial outburst.  Luminosity at
this stage, $\sim 10^{42}$ erg/s, is highly super-Eddington. More
exactly, the fireball emits radiation in the extraordinary mode, for
which radiation opacities are strongly suppressed by magnetic field,
and the fireball luminosity is in fact the Eddington luminosity for
this radiation \citep{Thompson1995}. However, extraordinary photons
with the energy $\PH{E}>40$ keV are split in the magnetosphere to
ordinary mode photons, for which the magnetized opacities are
comparable with the non-magnetized ones. Therefore any plasma in the
magnetosphere will be pushed upwards by the radiation from the
fireball. However, the magnetosphere could not become empty because it
should support electric currents generated during the initial outburst
when the the magnetosphere was not force-free. If the plasma density
falls below the value necessary to maintain the current, the electric
field (displacement current) arises and accelerates particles until
they produce electron-positron pairs
\citep{Thompson2002,Beloborodov2007}.

During the tail phase, the pairs are easily produced via
$\gamma-\gamma$ conversion of photons upscattered off relativistic
electrons. The photon density in the magnetosphere is $\PH{n}=
10^{25}L_{41}/\varepsilon_{50}$ cm$^{-3}$, where $L=10^{41}L_{41}$
erg/s is the luminosity at the photon energy
$\varepsilon=50\varepsilon_{50}$ keV. The scattering cross-section of
the ordinary photon is $\sigma=\sigma_T\sin^2\theta'$, where $\theta'$
is the angle between the photon and the magnetic field in the proper
electron frame. So the electron with the Lorentz factor
$\gamma=10/\varepsilon_{50}$ produces a photon with the energy
$\gamma^2\varepsilon=10mc^2/\varepsilon_{50}$ after passing a length
$l=(\sigma n)^{-1}=20\varepsilon_{50}/L_{41}$ cm. This photon is
easily converted into a pair by collision with a background photon. In
principle, electrons with lesser Lorentz factors could also produce
pair by upscattering photons with higher energies however, the density
of photons sharply decreases with the energy therefore one can expect
that the electron is accelerated to at least $\gamma>10$ before it
produce a pair.  After the first pairs are produced, they screen the
electric field and acceleration ceases. The accelerated electrons
rapidly loose their energy upscattering photons. As any ordinary
photon with the energy $>1$ MeV is converted into a pair when it
acquires an appropriate angle with the magnetic field, one can expect
that eventually any primary electron produces at least a few pairs,
the number of pairs being proportional to the energy the electron
acquires within the gap. An important point is that the number density
of the primary electrons is proportional to the current density,
$n_\mathrm{primary}=j/ec$, because there are no other particles in the gap.
Therefore one can expect that the total number of the pairs is
proportional to the current
\begin{equation}
\label{n_trough_kappa}
n=\kappa j/ec;
\end{equation}
where the multiplicity $\kappa$ is at least of the order of a few but
may be larger. In order to find the multiplicity one should solve
self-consistently electrodynamics of the gap together with the
particle dynamics and pair-production \citep{Beloborodov2007}, which
is beyond the scope of this letter

\end{document}